\documentclass[%
 reprint,
 amsmath,amssymb,
 aps,prx
]{revtex4-2}

\usepackage{graphicx}
\usepackage{dcolumn}
\usepackage{bm}

\begin{document}

\title{Randomness with constraints:  constructing minimal models for high-dimensional biology}

\author{Ilya Nemenman}
 \email{ilya.nemenman@emory.edu}
 \affiliation{Department of Physics, Department of Biology, Initiative in Theory and Modeling of Living Systems, Emory University, Atlanta, Georgia, 30322, USA}
\author{Pankaj Mehta}%
 \email{pankajm@bu.edu}
\affiliation{Department of Physics and Faculty of Computing and Data Science, Boston University, Boston, MA, 02215, USA}

\date{\today}

\begin{abstract}
Biologists and physicists have a rich tradition of modeling living systems with simple models composed of a few interacting components. Despite the remarkable success of this approach, it remains unclear how to use such finely tuned  models to study complex biological systems composed of numerous heterogeneous, interacting components. One possible strategy for taming this biological complexity is to embrace the idea that many biological behaviors we observe are ``typical'' and can be modeled using random systems that respect biologically-motivated constraints. Here, we review recent works showing how this approach can be used to make close connection with experiments in biological systems ranging from neuroscience to ecology and evolution and beyond. Collectively, these works suggest that the ``random-with-constraints'' paradigm represents a promising new modeling strategy for capturing experimentally observed dynamical and statistical features in high-dimensional biological data and provides a powerful minimal modeling philosophy for biology.
\end{abstract}

\maketitle


\section{Introduction}
Biology has a long and extremely successful tradition of modeling complex  phenomena using systems with just a few interacting components. In neuroscience, the Hodgkin–Huxley equations, a two‐channel model of sodium and potassium currents, have served for more than 70 years as the prototypical model for generating action potentials \cite{hodgkin1952quantitative}.  In ecology, Lotka and Volterra analyzed the dynamics of two interacting species (foxes and hares) to capture cyclical population changes, demonstrating the power of simple  models to elucidate complex ecological dynamics \cite{lotka1925elements,volterra1926variazioni}. Systems biology has similarly embraced network motifs, such as feedforward and feedback loops and other small circuits with three or four nodes, as the elemental modules of genetic and signaling networks \cite{milo2002network}.  Even at the level of physiology and medicine, small circuits have been a useful framework for understanding homeostasis and control \cite{karin2016dynamical}.  These paradigmatic examples follow the same logic: start with tractable circuits that yield intuitive functional insights, and put them together to build ever-larger systems.

\begin{figure*}[t]  
  \centering
  \includegraphics[width=\textwidth]{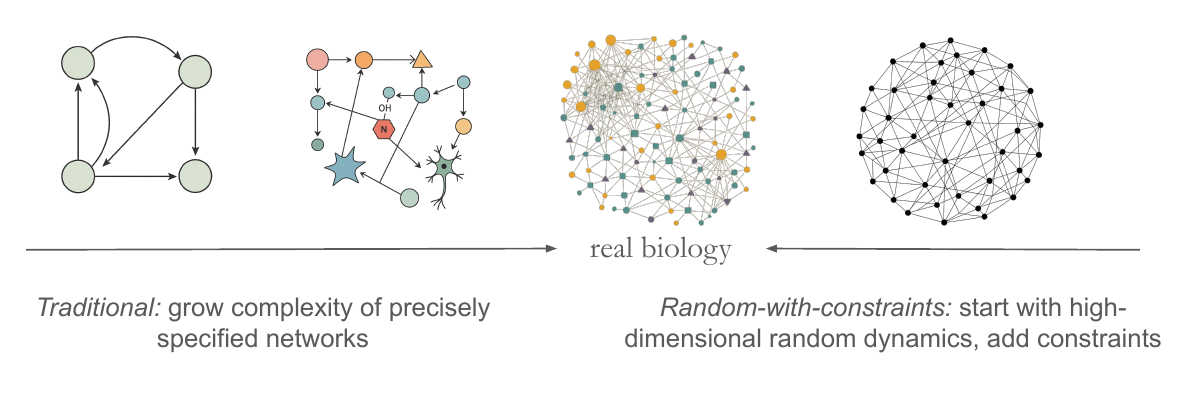}
  \caption{Traditional approaches to modeling in biology start with a precisely specified small system, which is then made more realistic by adding more elements and interactions. In contrast, the starting point in the  ``random-with-constraints'' approach is a high-dimensional random biological network. To make the model more realistic, one imposes more and more biologically-informed constraints. This latter approach is more suited for making connections with ``high-dimensional'' biological data. }
  \label{fig:logic}
\end{figure*}

Yet real biological systems are vastly more complex. Cortical microcircuits encompass thousands of neurons with heterogeneous cell types and connectivity; ecosystems comprise dozens to hundreds of species with a mix of global and local interactions mediated by the environment; cellular networks involve hundreds of genes, RNAs, and proteins forming dense networks of mass and information flow. Bottom‐up models that start with a handful of components often struggle to predict emergent behaviors when used in such high‐dimensional contexts. One potential approach to this challenge is to use high-throughput experiments to fit statistical models that capture interactions between components. Yet, even using the largest datasets currently available, it remains impossible to accurately fit interactions among all components, inevitably leading to many poorly constrained parameters. This highlights the deep tension in biological modeling between tractable minimality and complexity, see Fig.~\ref{fig:logic}.

Network science \cite{newman2018networks} has sought to bridge this gap by emphasizing structural constraints rather than detailed mechanisms.  Studies of static network topology use random-graph null models, e.g., the Erd\"os–R\'enyi graphs \cite{erdos1959on} or other model networks that preserve degree distributions \cite{barabasi1999emergence}, to ask whether observed features (e.g., clustering, motifs) exceed chance expectations.  For example, one can compare the frequency of a feedforward loop in a real gene network to its frequency in a randomized one to infer its functional importance \cite{milo2002network}.  However, these approaches remain largely structural: they tell us which patterns occur but not how dynamics on those patterns give rise to function.

One potential inspiration for overcoming these challenges comes from nuclear physics.  Heavy nuclei contain hundreds of protons and neutrons interacting by forces too complex to specify microscopically.  In the 1950s, Wigner’s surmise replaced the detailed nuclear Hamiltonian with an ensemble of random matrices obeying the same symmetry constraints (i.e., Hermitian).  Wigner and Dyson then calculated the expected level‐spacing distribution (not each energy level in each specific nucleus, but the statistics across all nuclei), predicting a characteristic level repulsion \cite{wigner1951statistical,dyson1962statistical}. This prediction matched experimental data remarkably well. This ensemble-based, statistical approach revealed universal properties of heavy nuclei without detailed mechanistic input, creating a baseline to which experimental measurements and theoretical calculations could be compared. At the heart of this approach was the realization that when systems become sufficiently complex, many properties of these systems become ``typical'' and hence can be modeled using random interactions with constraints. Then the deviations from the ``ideal'' embodied in the Wigner surmise naturally highlight the unique physics of particular isotopes and heavy nuclei.

We argue that a similar approach centered on typicality and constraints is ideally suited for taming the complexity of high-dimensional biological systems, see Fig.~\ref{fig:logic}. In this approach, one treats biophysical constraints as structural inputs to an ensemble of random networks and studies the dynamics such ensembles produce as a minimal quantitative model that represents the typical behavior of a high-dimensional biological system.  Rather than fitting every reaction rate, synaptic weight, or ecological interaction parameter, one specifies broad constraints---species classes, interaction types and symmetries, connectivity statistics, conservation laws, evolutionary constraints, or biophysical limits---and then draws interactions at random within those constraints.  By analyzing the emergent dynamics (e.g., structure and number of attractors, transients on the way to them) one uncovers generic functional behaviors that hold across the ensemble.  Just as Wigner showed that statistical spectra reveal core physics of complex nuclei, constrained random network ensembles can reveal principles of neural computation, ecological stability, cellular regulation, and physiological homeostasis, without committing to a single, fully detailed wiring diagram. In other words, they can reveal the {\em physics} of the underlying phenomena, with deviations from typicality indicating the likely presence of interesting {\em biology}.

This constrained random ensemble paradigm offers an alternative ``minimal model'' philosophy, complementing traditional approaches based on small circuits. It balances simplicity and realism, focuses on emergent dynamics rather than microscopic exactitude, and readily accommodates quantitative comparisons with high‐dimensional data.  In what follows, we provide some historical background for this approach and  survey recent empirical successes across biological domains, thus laying the groundwork for what we hope,  ``perhaps\dots too courageous[ly]'' \cite{wigner1957conference}, will become a powerful new paradigm for modeling complex living systems.

\section{Theoretical Background}
The idea to use large random networks to explore biology, and specifically dynamical processes in biology, treating complex, interconnected systems as statistical ensembles rather than as meticulously specified circuits, is certainly not new. Theorists have been doing this since the late 1960s, only about a decade after Wigner and Dyson. The biggest difference, compared to the more recent advances, is that they had to do it with very little experimental data to guide their thinking.  

In one of the earliest examples, Stuart Kauffman \cite{kauffman1969metabolic} proposed random Boolean networks as generic gene‐regulatory models, in which each gene’s on/off state is updated according to a randomly chosen Boolean function of its inputs. He used these ensembles to study how global dynamics (e.g., order, chaos, and canalization) emerge from generic wiring.  Shortly thereafter, Niels Jerne \cite{jerne1974towards} developed the idiotypic network theory of the immune system, in which  lymphocytes and antibodies form a vast regulatory network whose interactions are essentially random. Although mainly qualitative, this idea initiated thinking about immune memory and tolerance from a (random) network perspective, bringing methods of statistical physics to advance theory \cite{perelson1979theoretical,perelson1989immune,de1991size, parisi1990simple}.  

In community ecology, Robert May’s landmark  analysis showed that if one constructs a community matrix of species interactions with randomly drawn strengths, then the condition for linear stability predicts that increasing complexity (that is, community size) destabilizes equilibria, thus contradicting the prevailing intuition that biodiversity leads to resilience \cite{may1972will}.  Although May’s work was purely theoretical, it launched the stability-complexity debate and seeded countless extensions, where powerful methods of random matrix theory and statistical physics of disordered systems were used to explore the tradeoff for ever more constrained random interaction ensembles.

In molecular biophysics, Bernard Derrida’s random‐energy model \cite{derrida1980random} treated the energies of all microstates in a disordered system as independent random variables, yielding a solvable spin‐glass–like landscape with a phase transition into a ``frozen'' phase. This minimal  model provided the conceptual foundation for later spin‐glass approaches to protein folding. For example,  Bryngelson and Wolynes \cite{bryngelson1987spin} mapped protein folding onto a spin‐glass heteropolymer with random monomer interactions, deriving a mean‐field phase diagram in which a protein‐like funnel‐shaped landscape emerges from sufficiently unfrustrated random heteropolymer interactions.

Finally, in theoretical neuroscience, Daniel Amit and collaborators took inspiration from seminal papers by John Hopfield to develop models for memory storage and retrieval using attractor neural networks \cite{amit1989modeling}. Building on this work, Sompolinsky, Crisanti and Sommers \cite{sompolinsky1988chaos} introduced a continuous‐time neural network of many rate‐based neurons coupled by Gaussian random synaptic weights and solved its dynamics via dynamical mean‐field theory (DMFT), demonstrating emergence of high‐dimensional chaos.  Subsequent extensions by van Vreeswijk and Sompolinsky \cite{van1996chaos} constraining the network ensemble to balance random excitatory–inhibitory (E-I, or positive-negative) synaptic strength can produce irregular, intermittent, asynchronous states akin to cortical activity, again on the basis of statistical structure rather than detailed wiring.  Parenthetically, these methods then diffused back into community ecology, as we discuss below.
Finally, though outside of the context of living systems, such random neural networks even became useful artificial intelligence tools \cite{jaeger2001echo,maass2002real}.

\begin{figure*}[t]  
  \centering
  \includegraphics[width=\textwidth]{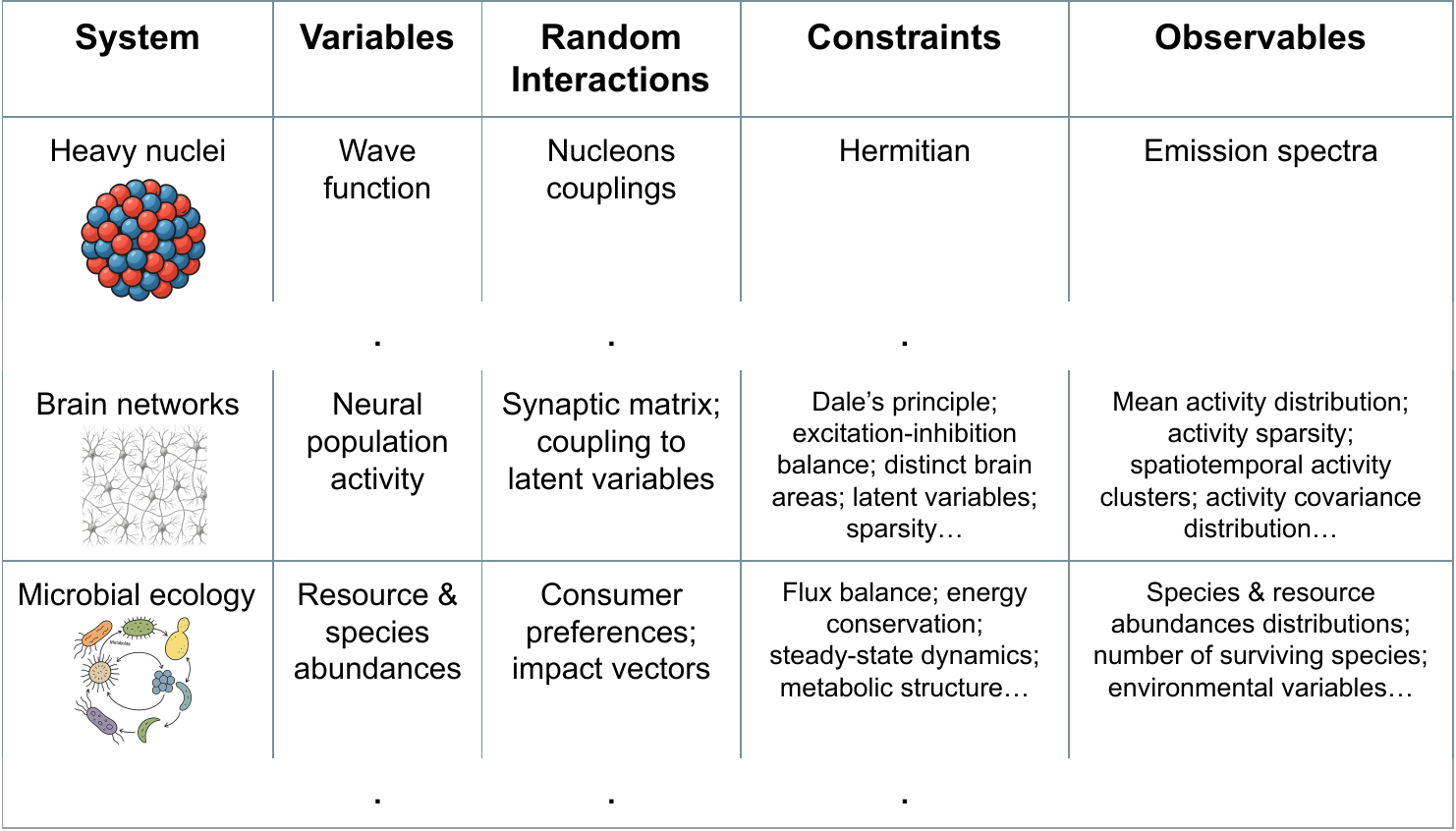}
  \caption{``Random-with-constraints'' approach started in nuclear physics. Applications in neuroscience and ecology, discussed in this perspective, work with different variables, predict different experimental observables, and, crucially, use different, more complex, constraints on ensembles of random interactions. Applications to different domains of biophysics and quantitative biology will require further variations of constraints. }
  \label{fig:table}
\end{figure*}

Throughout these fields, the common thread has been to sacrifice microscopic specificity in favor of ensemble statistics. One starts with specifying and constraining the ensemble of variables and interactions, e.g., excitatory and inhibitory neurons, or chemical species produced and degraded by biological species.  One then uses physics tools---replica theory, cavity methods, DMFT, numerical simulations---to extract generic behaviors (e.g., phase transitions, stability criteria, spectral properties) of ``typical'' networks under broad structural constraints.  In each case, early investigators recognized that data were insufficient to constrain all parameters, so they asked what could happen in a high-dimensional system, rather than what actually happens in a particular instance.  Only in recent years have experiments in neuroscience, microbial ecology, protein folding, and immunology begun to test these theoretical predictions quantitatively. Yet the rich legacy of random-network-with-constraints models provides a powerful and well-developed conceptual and mathematical toolkit. We argue that this toolkit should now be embraced as a minimal, quantitative paradigm for understanding complex biological dynamics (Fig.~\ref{fig:logic}), just as Wigner’s surmise revolutionized nuclear physics by focusing on the typical spectra of random Hamiltonians. Yet since living systems and their components are a lot more diverse than protons and neutrons, and interactions in biology rarely obey simple symmetry principles, the ensembles of random networks that we expect to be useful in biology are very different from those studied by nuclear physicists, Fig.~\ref{fig:table}.

\section{Challenging random network models by experiments}

\subsection{Neuroscience}

As the number of neurons one can record from simultaneously has steadily increased over the last decades, and the connectomics data exploded, large-scale fine-tuned circuit models have not kept up with assimilating these data. Instead, random network models have increasingly moved from purely theoretical constructs to tools that quantitatively account for neural data across scales and modalities, Fig.~\ref{fig:table}. For example, the dynamics of the original random neural network models with and without E-I balance \cite{sompolinsky1988chaos,van1996chaos} have recently been compared to the covariance spectrum of the neural activity in a larval zebra fish, and the agreement between the models and the theory was much better than one would expect in a null model \cite{hu2022spectrum}. Further,  a wealth of experimental evidence points to the E-I balance in the cortex being ``loose'', a signature of random balanced networks, and how this feature naturally yields nonlinear population responses observed in experiments \cite{ahmadian2021dynamical}. 

In a similar vein, we previously compared predictions of random balanced neural network models to high‐dimensional neural population recordings, demonstrating that, without any tuning, such networks generate single‐cell selectivity and correlation patterns observed among excitatory and inhibitory cells in the mouse cortex \cite{sederberg2020randomly}. Further, together  with \cite{huang2019circuit}, \cite{sederberg2020randomly} established that various excitation-inhibition balanced models create low-dimensional correlated activity in neural populations, with external signals modulating the ``importance'' of different activity modes. 

In all these cases, the ``random-with-constraints''  means existence of distinct neural types, with synapses of appropriate signs and strengths (the so-called {\em Dale's principle}). However, different experiments, such as recordings from the primate prefrontal cortex, may require different constraints to be modeled. One possibility is  setting a randomly connected network into the reservoir computing regime \cite{singer2013cortical,enel2016reservoir}. Alternatively, modeling random networks with connectivity patterns resembling connectivity within and across brain areas qualitatively predicts properties of information routing between brain areas \cite{clark2025structure} and quantitatively accounts for the observation that more diverse firing rates of neurons in a population correlate with the lower effective dimension of the  trial-to-trial co-variation \cite{tian2024firing}.

Constraining random networks to be bipartite, where one class of nodes corresponds to observable neurons and the other to slowly varying latent degrees of freedom (which can be other neurons, external stimuli, or emergent dynamical variables) similarly explains---sometimes with the precision of two significant digits---experimentally observed emergence of critical scalings in the neural clusters activity in the rodent hippocampus \cite{morrell2021latent,meshulam2019coarse}, as well as in dynamics of neural avalanches across many different experimental systems \cite{morrell2024neural}. Extending these ideas to a network of neurons with a similarity defined by their geometric distance (i.e., the Euclidean random matrix \cite{mezard1999spectra} coupling model) also explains the invariance to subsampling in the neural activity covariance spectrum in whole‐brain imaging of zebrafish  \cite{wang2025geometry}.

Even classic sensory circuits can be understood through randomness. For example, mouse V1 orientation tuning, long thought to require spatially offset ON and OFF inputs, can emerge from purely random thalamocortical convergence and recurrent interactions \cite{pattadkal2018emergent}. Similarly,  a ``randomly clustered'' recurrent neural network with inhibitory feedback  was proposed to explain experimental hippocampal place-cell data, including various statistics of place fields  and  spontaneous replay/preplay sequences.  Agreement with  measurements in rats on first-exposure to a maze was very good \cite{breffle2024intrinsic}. Similar spatially-embedded random recurrent neural network models with balanced excitation and inhibition  can   produce extended attractors \cite{natale2020precise} or distinctive ``Mexican-hat'' correlations among neural pairs in macaque V1, observed experimentally \cite{rosenbaum2017spatial}.

Another possible constraint on the structure of networks is sparsity, so that most entries in the synaptic connectivity matrix are zero. In this context, joint population activity of hundreds of neurons in the monkey visual cortex can be approximated with quantitative precision by a model inspired by sparse, random connectivity of real neural circuits \cite{maoz2020learning}. Similarly, optimal sparse random connectivity models can predict the degree of synaptic connectivity observed experimentally in  the cerebellar granule cells and {\em Drosophila} Kenyon cells \cite{litwin2017optimal}.

Together, these diverse examples argue forcefully for a paradigm shift in computational neuroscience: complementing hand-crafted minimal models with a handful of components, a different type of minimal models is emerging, which embraces ensembles of random interactions constrained only by broad biological priors, Fig.~\ref{fig:table}. 

\subsection{Ecological and evolutionary dynamics}

The last decade has also seen a resurgence of interest in using random ecosystems to understand community ecology, especially microbial ecosystems (see \cite{cui2024houches} for a recent pedagogical introduction).  The primary driver for this is the widespread availability of DNA-sequencing-based techniques that allow experimentalists to quantitatively measure the abundance of species present in a microbial ecosystem. This has resulted in a flood of large-scale datasets on both natural communities and carefully designed lab experiments \cite{turnbaugh2007human, lloyd2017strains,cheng2022design}. In  turn, the high-dimensional datasets have allowed experimentalists to identify interesting large-scale patterns (species abundance relationships, modularity, etc.) in microbial ecology that demand explanation \cite{marsland2020minimal}, Fig.~\ref{fig:table}. 

This experimental progress has been complemented by new theoretical methods for analyzing complex ecosystems with many resources and species. Traditionally, models in theoretical community ecology, such as Lotka-Volterra models and consumer resource models, have been formulated for small ecosystems with just a handful of species and resources. Recently, statistical physicists have extended them to the ``high-dimensional'' limit, using the tools originally from theoretical neuroscience \cite{sompolinsky1988chaos} to study large ecosystems \cite{fisher2014transition, bunin2017ecological,tikhonov2017collective,pearce2020stabilization,cui2024houches}. Since it is impossible to measure all ecological interactions experimentally, physicists have naturally adopted the random-with-constraints paradigm to ask if one can explain the patterns seen in large-scale ecological datasets.

This program has been remarkably successful. One insight has been that, just like physical systems, ecosystems can undergo phase transitions as a function of ecological parameters (i.e., from niche-like phase where species interact strongly to a  ``neutral'' phase where species are effectively non-interacting or from a steady-state to a dynamic phase) \cite{fisher2014transition, hu2022emergent}. This prediction motivated recent experiments, which demonstrated that the  phase transitions may occur in large-scale ecosystems \cite{hu2022emergent}.

These models have also highlighted the central role of metabolism and metabolic cross-feeding in shaping microbial ecosystems. Numerous works now show that large-scale experimental patterns seen in systems ranging from soil microbes, to oceans, to the gut microbiome emerge generically from considering random consumer resource models augmented with metabolic constraints \cite{goldford2018emergent, marsland2020minimal, dal2021resource, ho2022competition}, Fig.~\ref{fig:table}. At a slightly more coarse-grained level, a similar random-with-constraint approach has proved highly successful at explaining macroecological patterns across space and time \cite{grilli2020macroecological, shoemaker2025macroecological}. There are now even Python packages for simulating complex ecosystems using this approach to allow quick comparison with experiments \cite{marsland2020community,gao2023miasim}.

Similar ideas have also recently been explored in the context of evolution. A beautiful example is the recent work showing that experimentally observed global patterns of epistasis---gene interactions,  where one gene modifies the effect of another gene on a phenotype---can be reproduced by considering high-dimensional, random fitness functions \cite{reddy2021global}. These models also make specific quantitative predictions for the relationship between the magnitude of global epistasis and the stochastic effects of microscopic epistasis. These predictions were successfully verified by reanalyzing experimental data, suggesting that the random-with-constraints approach is also likely to be a fruitful paradigm for understanding complex evolutionary traits.

A handful of papers have also started to use random systems to understand experimental eco-evolutionary dynamics---systems such as microbial communities where ecology and evolution can occur on similar scales \cite{good2023eco,good2018adaptation,fisher2021inevitability, feng2025theory}. A key theme emerging from these works is that, in contrast to predictions from low-dimensional population genetic models, closely-related strains, including parents and mutants, can often co-exist due to ecological interactions. Such co-existence has been a puzzling feature of modern experimental datasets on the temporal dynamics of microbial ecosystems \cite{goyal2022interactions,good2023eco}. However,  ``random-with-constraints'' models of eco-evolution suggest that such co-existence is a generic feature of high-dimensional eco-evolutionary systems.

\subsection{Other fields}
Neuroscience and ecology are the fields we work in, and hence the primary focus of this perspective. Yet, in other fields of quantitative biology and physics of living systems, from soft matter physics to immunology, the emerging picture is similar: random-with-constraints network models are slowly morphing from being purely theoretical exercises into  becoming useful tools for understanding experimental data with quantitative precision. Here we offer just a few examples. 

In the context of systems biology, gene regulatory networks can be modeled by a system of ODEs with random weights, quite similarly to neural or ecological networks. Such models explain experimental effects of  microRNAs on the overall system stability \cite{chen2019gene} with approaches very similar to those used by May in ecology \cite{may1972will}. Further, the statistics of completion time distributions for some complex biochemical processes can be explained, at least semi-quantitatively, by considering chemical reactions as traversing random graphs of states \cite{bel2009simplicity}. 

In soft matter biophysics, bulk elasticity properties of collagen and other soft tissues are modeled very well by a random-fiber network model \cite{beroz2017physical}, which quantitatively fits the entire range of linear and nonlinear elasticity data, for both healthy and fibrotic-like tissues \cite{jansen2018role}. Similarly, random-network models of lung-alveolar wall mechanics reproduce stress-strain curves  for healthy vs fibrotic alveoli \cite{casey2021percolation}. These examples parallel uses of random, constrained network models in the physics of disordered systems, beyond living systems \cite{stanifer2018simple}.

In immunology, random bipartite antigen-receptor models have been used to understand the optimal organization the adaptive immune system \cite{mayer2015well}. Further, network models that respect that some immune cells activate and some suppress each other (similar to the E-I bipartite structure of neural networks) have been used to predict how changes in self-antigen concentrations may  result in an autoimmune response \cite{marsland2021tregs}. Experimental confirmations of these models are still wanting, though imminently possible with modern technologies. 

\section{Discussion}

Our brief survey demonstrates that the ``random-with-constraints'' paradigm, in which interaction matrices or couplings are drawn at random from biologically constrained ensembles, Fig.~\ref{fig:table}, consistently captures experimentally observed  dynamical and statistical features of living systems across many domains of science.  These models offer a minimal yet quantitatively accurate framework that bridges the gulf between small-circuit mechanistic models and unwieldy, fully-specified simulations and complex statistical models.  We therefore argue for the serious adoption of constrained random-network ensembles as a core modeling paradigm in quantitative biology.  By focusing on broad structural constraints  rather than every microscopic interaction, one can often achieve predictive, quantitative descriptions without overfitting or intractable parameterizations.

We believe that the next steps should include systematic comparisons, across multiple fields of quantitative biology, of dynamics predicted by different random ensembles  to experimental data. Doing so would achieve two goals. First, additional success of constrained random networks as minimal models for data would provide an even stronger justification for our hypothesis. Second, such additional examples will help us understand which structural constraints are useful for which types of data. Ideally, analyzing the errors that such models  make will allow for a better understanding of how to better formulate constraints in complex biological systems. By iterating this procedure, it should be possible to produce even more accurate---but still largely random---models.

A vital open question is why random-network models work so well. We need to understand whether there exist universal principles, akin to universality classes in statistical physics, that guarantee that some specific ensembles will reproduce core observables.  Equally important is developing predictive criteria for selecting the appropriate constraint set in advance: Can we infer, from preliminary data or system architecture, which structural features are essential for modeling a given phenomenon?  Answering these questions could transform random-ensemble modeling from an art into a science.

One intriguing possibility is that random networks may approximate biological systems effectively because living systems must operate in a world whose statistics are themselves well captured by random-feature models.  Indeed,  we  argued that the statistics of the natural world (specifically, small image patches) can be approximated exquisitely well by a sparse mixture of random linear latent features \cite{fleig2025hidden}.  In other words, the success of random network models may ultimately reflect an alignment between the statistical structure of the environment and the emergent dynamics of high-dimensional random systems. 

Another possibility is that the success of these models reflects the random nature of evolution.  It is plausible that evolutionary solutions to biological constraints in high-dimensional systems do not require fine tuning of all interactions. Instead, just as in modern machine learning models, evolution finds local minima that satisfy functional and evolutionary constraints with many flexible parameters and interactions that can vary considerably or even be pruned \cite{goaillard2021ion, marder2011multiple, frankle2018lottery}. If this is the case, then any random network that satisfies the relevant constraints should produce similar dynamics as the real biological system.

A final possibility is that even when non-specific interactions are weak, in systems with many interacting components, these non-specific interactions dominate the large-scale patterns one observes in high-throughput experiments. This was explicitly shown  in \cite{cui2021diverse} which analyzed an ecological consumer resource model and showed that even for highly-structured ecosystems, introducing a small amount of random interactions made large-scale properties of these structured complex ecosystems indistinguishable from random consumer resource models. 

Whatever the reason for their success, it is clear that models that combine randomness with biologically-motivated constraints work well at recapitulating patterns seen in experimental data across a wide variety of biological systems from neuroscience, to ecology and evolution, and beyond. It is clear that the ``random-with-constraints'' paradigm provides a powerful  minimal modeling philosophy for biology---or maybe for physics of living systems. This approach balances parsimony and realism, emphasizing emergent dynamical principles, and, crucially, yields quantitative agreement with experimental data. We suspect that in the next decade, we will have many more examples of the successes of this philosophy. 

\begin{acknowledgments}
Both authors are grateful to Emory University and an anonymous donor, who supported a 2019 workshop on ``What is theoretical biological physics in the age of quantitative biology and big data?'', where some of the ideas in this perspective originated. IN thanks Audrey Sederberg, who was instrumental in forming his thinking about the problem. IN was supported, in part, by the Simons Investigator program. PM would like to thank Wenping Cui, Joshua Goldford, Robert Marsland, and Jason Rocks for helping develop the ideas presented here. PM would also like to acknowledge support from NIH NIGMS R35GM119461 and the
Chan-Zuckerburg Initiative. We also thank Thierry Mora, Aleksandra Walczak, and Aneta Koseska for suggesting relevant references.
\end{acknowledgments}

\bibliography{RandomNets}

\end{document}